# A novel crystal polymorph of volborthite, $Cu_3V_2O_7(OH)_2 \cdot 2H_2O$

Hajime Ishikawa,[a*] Jun-ichi Yamaura,[a] Yoshihiko Okamoto,[a] Hiroyuki Yoshida,[b] Gøran J. Nilsen[a] and Zenji Hiroi[a]

[a]Institute for Solid State Physics, University of Tokyo, Kashiwanoha 5-1-5, Kashiwa 277-8581, Japan, and [b]National Institute for Materials Science, Sengen 1-2-1, Tsukuba 305-0047, Japan

A new polymorph of volborthite [tricopper(II) divanadium(V) heptaoxide dihydroxide dihydrate], $Cu_3V_2O_7(OH)_2 \cdot 2H_2O$, has been discovered in a single crystal prepared by hydrothermal synthesis. X-ray analysis reveals that the monoclinic structure has the space group $C2/c$ at room temperature, which is different from that of the previously reported $C2/m$ structure. Both structures have $Cu_3O_6(OH)_2$ layers composed of edge-sharing $CuO_4(OH)_2$ octahedra, with $V_2O_7$ pillars and water molecules between the layers. The Cu atoms occupy two and three independent crystallographic sites in the $C2/m$ and $C2/c$ structures, respectively, likely giving rise to different magnetic interactions between CuII spins in the kagome lattices embedded in the $Cu_3O_6(OH)_2$ layers.

Kagome antiferromagnets have been studied extensively in a search for exotic magnetic phenomena. It is believed that an unknown ground state, such as a spin liquid, can be realized instead of the conventional Néel order, due to geometric frustration of spins residing on triangle-based lattices. Several Cu minerals comprising kagome lattices formed by $Cu^{II}$ ions have been targeted as model compounds for the spin-1/2 kagome antiferromagnet: herbertsmithite $ZnCu_3(OH)_6Cl_2$ (Shores *et al.*, 2005), volborthite $Cu_3V_2O_7(OH)_2 \cdot 2H_2O$ (Hiroi *et al.*, 2001), vesignieite $BaCu_3V_2O_8(OH)_2$ (Okamoto *et al.*, 2009), haydeeite $Cu_3Mg(OH)_6Cl_2$ (Colman *et al.*, 2010) and kapellasite $Cu_3Zn(OH)_6Cl_2$ (Colman *et al.*, 2008). However, the true ground state of the spin-1/2 kagome antiferromagnet is still a mystery because of experimental obstacles posed by real compounds; distortion, defects and other deviations from the ideal kagome model tend to hinder the characterization of low-temperature properties. Volborthite, a mineral found in nature as yellow–green crystals of sub-millimetre size, has been known since the 18th century. Its chemical and physical properties were described by Guillemin (1956). The compound has recently been studied as a candidate for the spin-1/2 kagome antiferromagnet (Hiroi *et al.*, 2001). A high-quality powder sample of volborthite was synthesized hydrothermally and found to exhibit anomalous magnetic transitions at temperatures near 1 K (Bert *et al*., 2005; Yoshida, Okamoto *et al*., 2009; Yoshida, Takigawa *et al*., 2009).

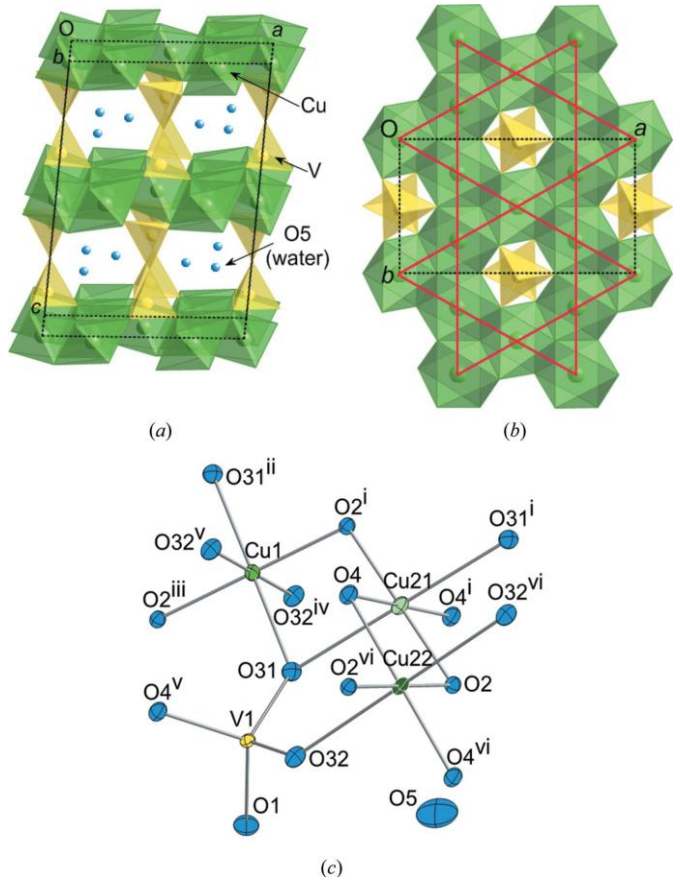

Figure 1

(a) Axial view, (b) plane view and (c) displacement ellipsoid plot, including the asymmetric unit, of the $C2/c$ phase of volborthite. Octahedral and tetrahedral polyhedra coordinated to Cu and V atoms, respectively, are displayed. $CuO_6$ octahedra form kagome layers separated by $V_2O_7$ pillars and solvent water molecules. The layer contains a distorted kagome lattice made up of $Cu^{II}$ ions, represented by thick lines in (b). Displacement ellipsoids are drawn at the 50% probability level in (c). [Symmetry codes: (i) -$x$+1/2, -$y$+1/2, -$z$; (ii) -$x$, -$y$, -$z$; (iii) $x$-1/2, $y$-1/2, $z$; (iv) $x$, $y$-1, $z$; (v) -$x$, -$y$+1, -$z$; (vi) -$x$+1/2, -$y$+3/2, -$z$.]

The crystal structure of volborthite at room temperature has been examined by several groups and still remains controversial. Leonardsen & Petersen (1974) described a monoclinic unit cell with $a$ = 10.604, $b$ = 5.879 and $c$ = 7.202 Å, and $β$ = 94.81°. Kashaev & Vasil'ev (1974) reported the space group $C2/c$ or $Cc$. Later, Basso et al. (1988) reported another monoclinic structure in the space group $C2/m$ by X-ray diffraction (XRD) measurements on a natural single crystal; this had the same unit cell as that observed by Leonardsen & Petersen (1974). A similar crystal structure in the space group $C2/m$ and the same unit cell was found by Lafontaine *et al.* (1990) through X-ray and neutron diffraction measurements using a synthetic powder sample. However, Kashaev *et al*. (2008) gave a different monoclinic structural model in the space group $Ia$. Thus, the crystal system is monoclinic in all cases, but it seems that more than one polymorph exists at room temperature. Very recently, Yoshida *et al*. (2012) successfully prepared a single crystal of sub-millimetre size and found a first-order structural phase transition at 310 K from a *C*2/*m* phase at high temperature to an *I*2/*a* phase at low temperature, as characterized by X-ray diffraction. The previous discrepancies regarding the structure may be partly related to the presence of this transition near room temperature. Although there are minor differences between these crystal structures, they commonly comprise $Cu_3O_6(OH)_2$ layers built of edge-sharing $CuO_4(OH)_2$ octahedra, which are separated by $V_2O_7$ pillars and unligated water molecules, as depicted in Fig. 1(a). In the $Cu_3O_6(OH)_2$ layer, $Cu^{II}$ ions form distorted kagome lattices and $V^V$ ions are located above and below the centres of the hexagons of the kagome lattice (Fig. 1b). To date, the magnetic properties of volborthite have been discussed on the basis of the *C*2/*m* structure reported by Lafontaine *et al*.

In this study, we report a new polymorph of volborthite in a synthetic single crystal of millimetre size, probably of higher quality than those studied previously. This sample has another monoclinic structure, in the space group *C*2/*c*, at room temperature, further illustrating the richness of the crystal chemistry of volborthite.

The *C*2/*c* structure found in the present study is a 2*c* superstructure of the *C*2/*m* system (Basso *et al*., 1988; Lafontaine *et al*., 1990). We observed 1543 supercell reflections that would not have integer indices for the *C*2/*m* cell, out of a total of 4212 reflections found for our cell in *C*2/*c*. The average $I$ and $I/\sigma(I)$ are 12752.3 and 43.2, respectively, for the principal reflections, and 661.0 and 21.8, respectively, for the supercell reflections. This structure may be identical to the *C*2/*c* or *Cc* structure given by Kashaev & Vasil'ev (1974), for which atomic coordinates are not available. Since there is no objective reason to question the veracity of the *C*2/*m* phase reported by Basso *et al.* (1988) and Lafontaine *et al.* (1990), we conclude that at least two polymorphs of volborthite exist at room temperature.

By way of comparison, there are two and three crystallographic sites for Cu in the *C*2/*m* and *C*2/*c* structures, respectively, as depicted in Figs. 1(c) and 2. In *C*2/*m*, atoms Cu1 and Cu2 occupy the 2*a* and 4*e* positions with site symmetries 2/*m* and -1, respectively. In contrast, in the present *C*2/*c* phase, atom Cu1 has lower site symmetry with -1 at the 4*a* position, and atom Cu2 splits into Cu21 and Cu22 at the 4*c* and 4*d* positions, both with inversion symmetry. Irrespective of these differences at the Cu sites, the kagome lattices in both structures consist of isosceles triangles formed by Cu1 and either Cu2 (*C*2/*m*) or Cu21/Cu22 (*C*2/*c*).

The two structures clearly differ in the environment of the Cu1 sites. The Cu—O bond lengths are compared in Table 1. In the *C*2/*m* structure, atom Cu1 is coordinated by six oxide ligands, with two short Cu1—O2 and four long Cu1—O3 bonds. (The 'long' bonds are intermediate in length between what is commonly found for short and long Cu—*L* distances in a Jahn–Teller compound; see below for an alternative explanation of this geometry.) In contrast, in the *C*2/*c* structure, there are four short bonds (two Cu1—O2 and two Cu1—O31) and two long bonds (Cu1—O32). The difference between the short and long bonds is large in the *C*2/*c* structure. In contrast, the coordination environments around Cu2 (*C*2/*m*) and Cu21/ Cu22 (*C*2/*c*) are nearly equal; in both structures, there are four short bonds (two Cu2—O2 and two Cu2—O4 in the *C*2/*m* structure, and two Cu21/Cu22—O2 and two Cu21/Cu22—O4 in the *C*2/*c* structure) and two long bonds (two Cu2—O3 and two Cu21/Cu22—O31/O32, respectively).

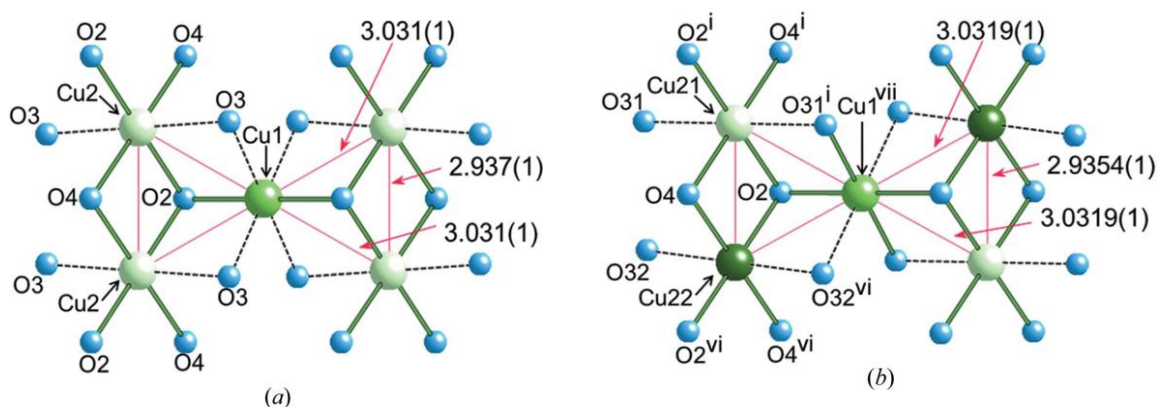

Figure 2

The coordination around the Cu atoms in (a) the *C*2/*m* structure (Basso *et al.*, 1988; Lafontaine *et al.*, 1990) and (b) the *C*2/*c* structure. Thin solid lines indicate Cu—Cu bonds, with lengths in Å. Short and long Cu—O bonds are displayed by thick solid lines and thin broken lines, respectively. [Symmetry codes: (i) -*x*+1/2, -*y*+1/2, -*z*; (vi) -*x*+1/2, -*y*+3/2, -*z*; (vii) *x*+1/2, *y*+1/2, *z*; generic atom labels are used in (a).]

Table 1

Cu—O bond lengths (Å) in the *C*2/*c* and *C*2/*m* volborthite structures.

| *C*2/*c* (this work) | *C*2/*m* (Lafontaine *et al.*) |
|---|---|
| Cu1—O2$^i$ 1.941 (3) | Cu1—O2 1.906 (10) |
| Cu1—O31 1.992 (2) | Cu1—O3 2.159 (7) |
| Cu1—O32$^{iv}$ 2.353 (2) | Cu2—O2 1.901 (6) |
| Cu21—O2 1.912 (2) | Cu2—O3 2.380 (8) |
| Cu21—O31 2.370 (3) | Cu2—O4 2.049 (15) |
| Cu21—O4 2.050 (2) | |
| Cu22—O22 1.923 (2) | |
| Cu22—O32 2.467 (3) | |
| Cu22—O4 2.005 (2) | |

Symmetry codes: (i) -*x* + 1/2, -*y* + 1/2, -*z*, (iv) *x*, *y*-1, *z*,

The details of the Jahn–Teller distortion of an octahedron about $Cu^{II}$ are important for understanding the magnetic properties of copper minerals, because the distortion determines the orbital state of the $Cu^{II}$ ion, and thus the magnetic interactions between neighbouring Cu spins. For an octahedron consisting of two short and four long Cu—O bonds, i.e. (2+4) coordination, an unpaired electron should occupy the $d_{z^2}$ orbital, while the $d_{x^2-y^2}$ orbital is occupied in the case of four short and two long bonds, i.e. (4+2) coordination. At the Cu1 site of the $C2/m$ structure, the $d_{z^2}$ orbital is apparently selected (but see below), while the $d_{x^2-y^2}$ orbital is occupied by an unpaired electron in the $C2/c$ structure. Such a difference in the orbital occupancies must give rise to a substantial difference in magnetic interactions between Cu spins in the kagome lattice. In particular, the magnetic interactions via Cu—O superexchange pathways between Cu1 and Cu2 spins are identical in the $C2/m$ structure, while those between Cu1—Cu21 and Cu1—Cu22 must be different in the $C2/c$ structure. It would be interesting to study the influence of this distortion on the magnetic properties of volborthite.

We have also carried out structural analyses at low temperatures and found a transition at 290 K to an $I2/a$ structure, which will be reported elsewhere. This lowtemperature structure is essentially the same as the $I2/a$ structure reported by Yoshida *et al.* (2012). Thus, volborthite takes the $I2/a$ structure at low temperatures, but there are two polymorphs in the vicinity of room temperature. This may be the main reason for the discrepancies among previous reports of the crystal structure at room temperature. The thermal history of a crystal might result in different structures, because the first-order transition to the $I2/a$ structure occurs with a thermal hysteresis of $\Delta T = 10$ K at around room temperature (296 K upon cooling and 306 K upon heating for Yoshida's crystal). In addition, another factor that may influence the structure is the presence of impurities or nonstoichiometry; a natural crystal contains some proportion of alien elements, such as other transition metals at the Cu or V site, while a synthetic crystal assumes no contamination.

Burns & Hawthorne (1996) point out that the (2+4) coordination for Cu1 in the $C2/m$ structure of volborthite can be attributed to the dynamic Jahn–Teller effect rather than to static order. In other words, two orthogonal (4+2) coordinations are dynamically swapped with each other, so that an apparent (2+4) coordination is observed in the average structure. If so, the present $C2/c$ and the low-temperature $I2/a$ structures could be considered as frozen states with one of two (4+2) coordinations being selected. On the other hand, Yoshida *et al.* suggest that the structural transition from $C2/m$ to $I2/a$ is related to an order–disorder transition involving the water molecules between the kagome layers. It is likely that ordering of these water molecules occurs in more than one way and also causes the structural differences between the $C2/m$ and $C2/c$ phases near room temperature. Alternatively, a subtle difference in the amount of water present could in principle result in two different structures. Further experiments are in progress to examine these possibilities.

Experimental

CuO (99.9%; 0.5727 g, 7.200 mmol) and $V_2O_5$ (99.99%; 0.4365 g, 2.400 mmol) powders were mixed (Cu:V 3:2 stoichiometric ratio) and, together with 1% nitric acid (15 ml), were placed in a Teflon container which was in turn placed in a stainless steel vessel. The vessel was sealed, heated to 443 K for 10 d and furnace-cooled to room temperature. Transparent yellow–green single crystals of volborthite were obtained, along with a polycrystalline powder. The crystals are arrow-head shaped and typically 1 ×0.5×0.05 mm in size. Several fractured crystals were used for single-crystal XRD measurements, all of which gave the same monoclinic $C2/c$ structure. Atomic positions were determined by the charge-flipping method. Occupational deficiency was not observed for the interstitial water molecules. Atom H2 was assumed to be present as part of an OH group (with O2), as found in the neutron diffraction refinement of the $C2/m$ structure by Lafontaine *et al*. Atom H2 was placed at a calculated position and refined as riding. The water H atoms could not be located.

Crystal data

$Cu_3V_2O_7(OH)_2 \cdot 2H_2O$, $M_r$ = 474.55, Monoclinic, $C2/c$
$a$ = 10.6118 (4) Å, $b$ = 5.8708 (2) Å, $c$ = 14.4181 (6) Å, $\beta$ = 95.029 (1) °, $V$ = 894.79 (6) Å$^3$, $Z$ = 4, Mo $K\alpha$ radiation, $\mu$ = 9.08 mm$^{-1}$, $T$ = 293 K, 0.11×0.06×0.02 mm

Data collection

Bruker APEX CCD area-detector diffractometer, Absorption correction: multi-scan (SADABS; Bruker, 2001), $T_{min}$ = 0.689, $T_{max}$ = 0.901, 8424 measured reflections 1030 independent reflections 954 reflections with $I > 2\sigma(I)$, $R_{int}$ = 0.026

Refinement

$R[F^2 > 2\sigma(F^2)]$ = 0.025 $wR(F^2)$ = 0.085, $S$ = 1.46, 1030 reflections, 79 parameters, H-atom parameters constrained, $\Delta\rho_{max}$ = 0.79 e Å$^{-3}$, $\Delta\rho_{min}$ = -0.78 e Å$^{-3}$

Data collection: APEX (Bruker, 2000); cell refinement: SAINT (Bruker, 2000); data reduction: SAINT; program(s) used to solve structure: SUPERFLIP (Palatinus & Chapuis, 2007); program(s) used to refine structure: SHELXL97 (Sheldrick, 2008); molecular graphics: CrystalMaker (Palmer, 2005); software used to prepare material for publication: SHELXL97.

This work was partly supported by a Grant-in-Aid for Scientific Research on Priority

Areas 'Novel States of Matter Induced by Frustration' (grant No. 19052003) provided by the Ministry of Education, Culture, Sports, Science and Technology, Japan.